# Strain energy enhanced room-temperature magnetocaloric effect in second-order magnetic transition materials

*Xiaohe Liu, Ping Song[*], Sen Yao, Yuhao Lei, Ling Yang, Shenxiang Du, Yiran Deng and Defeng Guo*

X. Liu, P. Song, S. Yao, Y. Lei, L. Yang, S. Du, Y. Deng, D. Guo

Center for Extreme Deformation Research, State Key Laboratory of Metastable Materials Science and Technology

Key Laboratory for Microstructural Material Physics of Hebei Province, School of Science

Yanshan University

Qinhuangdao 066004, P. R. China

Email: psong@ysu.edu.cn

**Abstract**

Large magnetic entropy change ($\Delta S_M$) can realize a prominent heat transformation under the magnetic field and directly strengthen the efficacy of the magnetocaloric effect, which provides a pioneering environmentally friendly solid-state strategy to improve refrigeration capacities and efficiencies. The second-order magnetic transition (SOMT) materials have broader $\Delta S_M$ peaks without thermal hysteresis compared with most first-order magnetic transition materials, making them highly attractive in magnetic refrigeration, especially in the room temperature range. Here, we report a significant enhancement of $\Delta S_M$ at room temperature in single-crystal $Mn_5Ge_3$. In this SOMT system, we realize a 60 % improvement of $-\Delta S_M^{max}$ from 3.5 J/kg·K to 5.6 J/kg·K at $T$ = 300 K. This considerable enhancement of $\Delta S_M$ is achieved by intentionally introducing strain energy through high-pressure constrained deformation. Both experimental results and Monte Carlo simulations demonstrate that the enhancement of $\Delta S_M$ originates from the microscopic strain and lattice deformation induced by strain energy after deformation. This strain energy will reconstruct the energy landscape of this ferromagnetic system and enhance magnetization, resulting in a giant intensity of magnetocaloric responses. Our findings provide an approach to increase magnetic entropy change and may give fresh ideas for exploring advanced magnetocaloric materials.

# 1. Introduction

The magnetocaloric effect (MCE), characterized by isothermal magnetic entropy change or adiabatic temperature change of magnetic materials under the application of an external magnetic field, provides a pioneering environmentally friendly solid-state strategy to improve refrigeration capacities and efficiencies.[1–4] Magnetic materials exhibiting a large MCE have been widely explored in the vast family of pure elements and compounds, ranging from rare earth materials to transition metal alloys, from bulk samples to nanoparticles, from single-phase materials to multiphase compounds, as reviewed in Ref. [5–7]. Magnetocaloric materials can be classified into first-order magnetic transition (FOMT) materials and second-order magnetic transition (SOMT) materials according to their phase transition order.[8,9] The FOMT materials often exhibit the advantage of a significant magnetic entropy change ($\Delta S_M$), but the accompanying thermal hysteresis and even the crystal lattice discontinuity will lead to undesired energy loss and physical instability during thermomagnetic cycling.[10]

In contrast, the SOMT materials have broader $\Delta S_M$ peaks without thermal hysteresis compared with those of most FOMT materials in the same operating temperature range, making them highly attractive in magnetic refrigeration, especially in the room temperature range.[11] The only fly in the ointment is that the SOMT materials usually possess low values of $\Delta S_M$ at operating temperatures. To date, researchers are still making efforts to explore the new SOMT materials with large $\Delta S_M$ mainly by component optimization and microstructure modification.[10,12–17] However, improving the inherent shortage of magnetocaloric responses at room temperature is still one of the enormous challenges for SOMT materials.

In this paper, we report a strategy to increase the magnetic entropy change of SOMT materials by intentionally enhancing the intensity of magnetization with high-pressure constrained deformation. The power of this anisotropic severe plastic deformation technique has also been verified in constructing high-performance permanent magnetic materials and antiferromagnetic materials.[18–27] Recently, this technique has achieved a breakthrough in the fast fabrication of multifunctional ferromagnets and led to the discovery of a class of multifunctional high-temperature ferromagnetic materials with an exceptional combination of high energy product, large electrical resistivity, and superior thermal stability of coercivity.[18] To demonstrate the efficacy of this strategy in the field of magnetocaloric materials, we select single crystal $Mn_5Ge_3$ as the model material. In this SOMT system, Mn and Ge atoms crystallize in the hexagonal space group $P6_3/mcm$ with Mn occupying two different Wyckoff positions of 4$d$ and 6$g$ sites and Ge occupying 6$g$ sites,[28] as shown in Figure S1. In this configuration, two different magnetic moments of Mn atoms align parallel to the hexagonal $c$-axis, forming a collinear ferromagnetic structure with a Curie temperature near 296 K [Figure S1].[11] The applied high-pressure constrained deformation will reconstruct the energy landscape of this ferromagnetic system by introducing strain energy. The additional strain energy will generate an enhancement in magnetization. As a result, we realize a giant intensity of magnetocaloric responses ($-\Delta S_M^{max}$ = 5.6 J/kg·K, $\Delta H$ = 1.6 T) at room temperature. This result is comparable to the value of Gd ($-\Delta S_M^{max}$ = 6.0 J/kg·K, $\Delta H$ = 2.0 T), which has been used as a benchmark material in magnetic refrigerator prototypes.[29] Our findings provide an approach to increase magnetic entropy change and may give fresh ideas for exploring advanced magnetocaloric materials.

## 2. Results and Discussion

According to Maxwell's relation, $(\partial S/\partial H)_T = (\partial M/\partial T)_H$, the magnetic entropy change ($\Delta S_M$) can be obtained as[1]

$$\Delta S_M = \int_0^H \frac{\partial M}{\partial T} dH,  \tag{1}$$

where $M$ is the magnetization, $T$ is the temperature, and $H$ is the external magnetic field. Equation (1) indicates that the increase in the magnitude of $M$ will result in a corresponding enlargement of $\Delta S_M$. Therefore, to achieve a significant $\Delta S_M$, we can design experimental strategies to increase the value of $M$. Our previous work demonstrates a feasible way to increase magnetization by introducing strain energy.[26,30] In this way, we may also achieve a large $\Delta S_M$ by realizing the increase of $M$ via the introduction of strain energy in the SOMT materials.

Figure 1 shows the concept of strain energy to enhance the magnetic entropy change in $Mn_5Ge_3$. According to the ferromagnetic thermodynamics of the Ising model, the magnetization ($M$) is a derivative of the total energy ($E$) of the external magnetic field ($H$) as[31]

$$M = \frac{\partial E}{\partial H}. \tag{2}$$

For the regular $Mn_5Ge_3$ crystal, the total energy $E$ includes exchange energy $E_1$ and Zeeman energy $E_2$ [Figure 1(a)]. When we introduce a lattice deformation, as shown in Figure 1(b), an additional strain energy ($E_U$) will arise from the lattice deformation. This additional $E_U$ will result in an increase in magnetization according to Equation (2) [Figure 1(c)]. As a result, we can realize a rise in $\Delta S_M$ [Figure 1(d)].

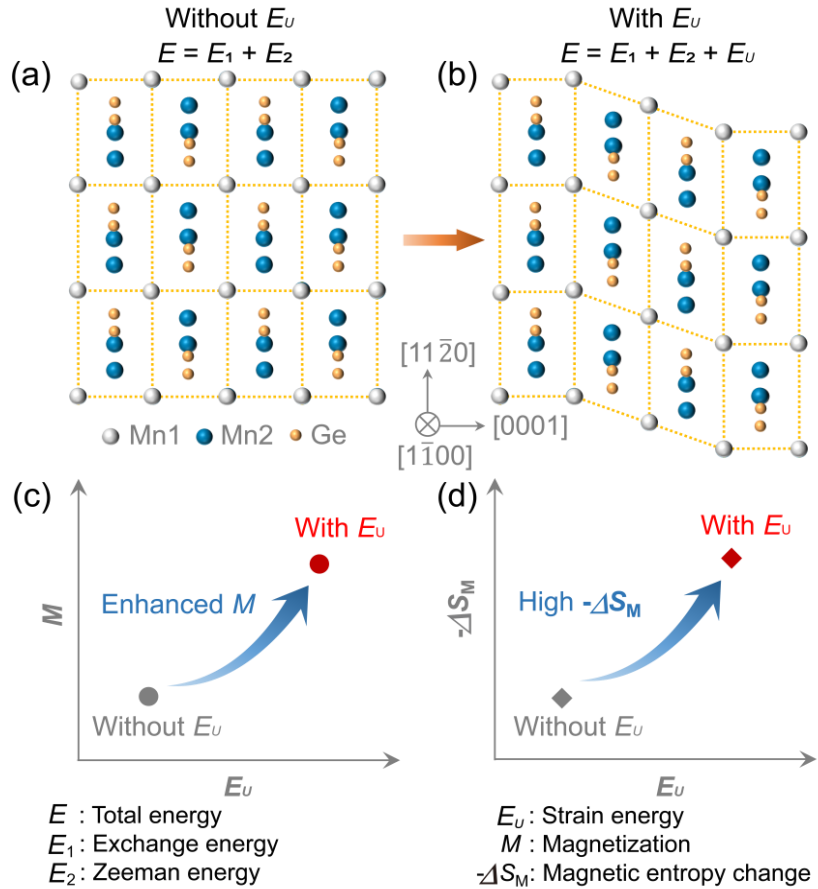

**Figure 1**. Concept of strain energy to enhance the magnetic entropy change in $Mn_5Ge_3$. Schematic diagram of the atomic configuration (a) without strain energy and (b) with strain energy on the ($1\bar{1}00$) plane. Schematic representation of the impact of strain energy on (c) magnetization and (d) magnetic entropy change.

Based on this idea, we design and experimentally introduce the strain energy in Mn$_5$Ge$_3$ by high-pressure constrained deformation. We first synthesize the Mn$_5$Ge$_3$ single crystals using the bismuth-flux method. The obtained Mn$_5$Ge$_3$ single crystals are in the form of a hexagonal prism, as shown in Figure S1. Figure S2 sketches the process of high-pressure constrained deformation for the Mn$_5$Ge$_3$ single crystals. The stress of deformation ($\sigma$) varies from 0 to 1.24 GPa, and the corresponding macro-strain ($\varepsilon$) ranges from 0 to 28 %. The morphology of the deformed Mn$_5$Ge$_3$ single crystals is shown in Figure S3. The energy-dispersive X-ray (EDX) measurement reveals a chemical composition of Mn$_{5.35}$Ge$_{2.84}$ for the as-prepared sample and Mn$_{5.35}$Ge$_{2.75}$ for the deformed sample (Figure S4). We refer to our samples as Mn$_5$Ge$_3$ in the following.

Figure 2(a) shows the XRD patterns of Mn$_5$Ge$_3$ at various values of $\sigma$. There are no extra diffraction peaks except $\{10\bar{1}0\}$ lattice planes that appear in the XRD patterns when $\sigma \leqslant 0.59$ GPa, indicating the single-crystal stability of Mn$_5$Ge$_3$ during the deformation process. With the increase in $\sigma$, the diffraction peaks of some other lattice planes and a new phase of Mn$_5$Ge$_2$ begin to appear at $\sigma = 0.75$ GPa. When $\sigma = 1.24$ GPa, the single-crystal sample shows an obvious poly-crystallization. By fitting the XRD patterns, we achieve the microscopic strain ($\varepsilon'$) of the deformed samples at different $\sigma$. As shown in Figure 2(b), the value of $\varepsilon'$ has a monotonic increase with the increase of $\sigma$. The strain energy density ($u$), $\varepsilon'$, and $\sigma$ are related as $u = \varepsilon'\sigma$[32]. The calculated value of $u$ has the same monotonically increasing trend as the increase in $\sigma$ [Figure 2(c)]. The result of $\sigma$ dependence of $\varepsilon'$ and $u$ indicates that strain energy is introduced into the sample by deformation.

To investigate the effect of the deformation process on the essence of the samples, we characterize the microstructure of Mn$_5$Ge$_3$ before and after deformation. Figures 2(d - f) show the transmission electron microscopy (TEM) image and selected area electron diffraction (SAED) pattern of the as-prepared Mn$_5$Ge$_3$ single crystals. The as-prepared single-crystal samples have a regular lattice arrangement without defects. In contrast, the deformed sample at $\sigma = 0.75$ GPa shows a prominent lattice deformation as marked by dash lines in Figure 2(g), although the SAED pattern shows a single-crystal characteristic [Figure 2(h)]. The high-resolution TEM (HRTEM) image of the deformed sample shows that the lattice planes of (11$\bar{2}$0) have apparent slip along the direction parallel to the stress [Figure 2(i)]. The prominent lattice deformation of the deformed samples may be facilitated by the enhanced strain energy under high stress and strain.[19] It should be noted that the HRTEM cross-section image of the deformed sample is not entirely along the pressure direction ([10$\bar{1}$0] crystal direction). This is due to the deviation caused by the limitation of experimental conditions in sample preparation.

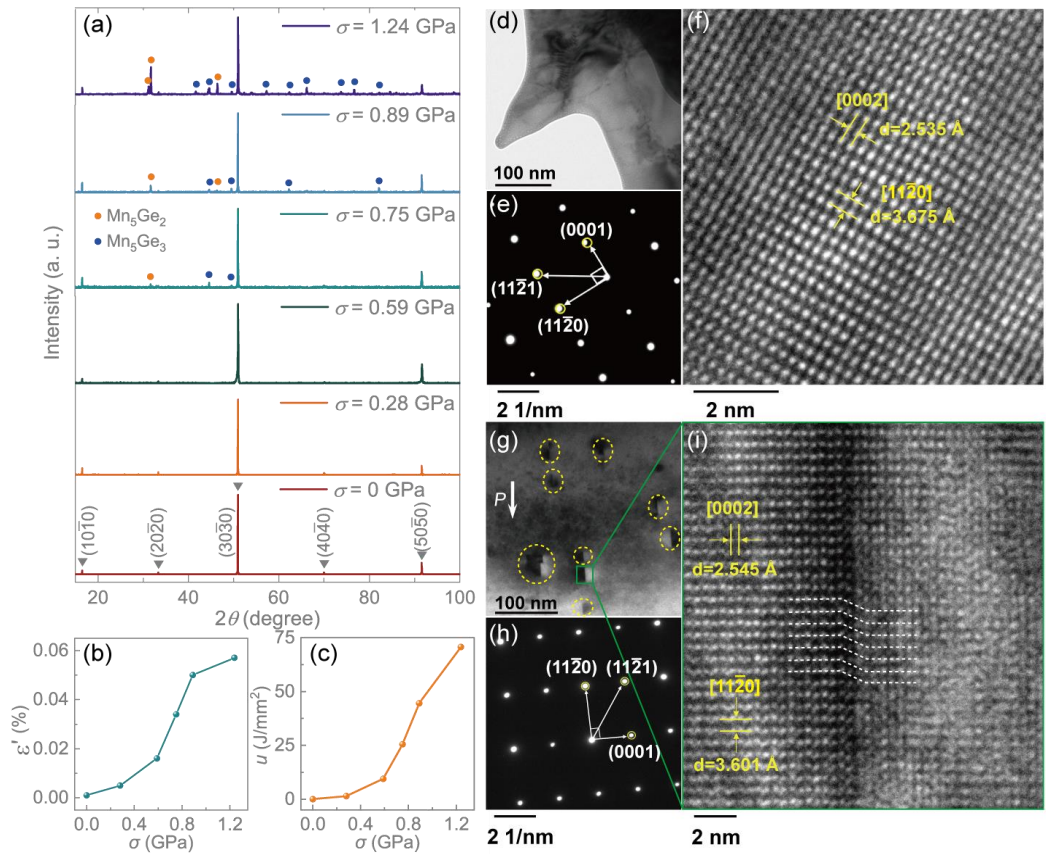

**Figure 2**. Crystal structure and microstructure of $Mn_5Ge_3$ single crystals before and after deformation. (a) XRD patterns with different stress ($\sigma$) measured parallel to [10$\bar{1}$0] direction. (b) The microscopic strain ($\varepsilon'$) fitted by XRD patterns and (c) the corresponding strain energy density ($u$) after the deformation at different $\sigma$. (d, e) TEM image and SAED pattern of the as-prepared $Mn_5Ge_3$. (f) High-resolution TEM (HRTEM) image of the as-prepared $Mn_5Ge_3$. (g, h) TEM image and SAED pattern of the cross-section along the stress direction at $\sigma$ = 0.75 GPa. (i) HRTEM image of the marked area in panel (g). White dash lines are to guide the eye to the lattice planes.

Figure 3 shows the results of $\Delta S_M$ modulated by deformation stress $\sigma$. As a representative plot, the magnetization isotherms of $Mn_5Ge_3$ at $\sigma$ = 0 GPa and $\sigma$ = 0.75

GPa are shown in Figure 3(a) and (b), and the results under other stresses are shown in Figure S5. From the magnetization isotherms, we calculate the corresponding values of $-\Delta S_M$ as a function of temperature and field changes. The results are presented in Figure 3(c) and (d), and the results under other stresses are shown in Figure S6. The applied external magnetic field $H$ parallels the [0001] direction for all the measurements, as indicated in Figure 3(a). The value of maximum magnetic entropy change ($-\Delta S_M^{max}$ = 3.5 J/kg·K) at $\sigma$ = 0 GPa for $\Delta H$ = 1.6 T is close to that reported in Ref. [11]. It is worth noting that the value of $-\Delta S_M^{max}$ at $\sigma$ = 0.75 GPa can reach 5.6 J/kg·K for $\Delta H$ = 1.6 T.

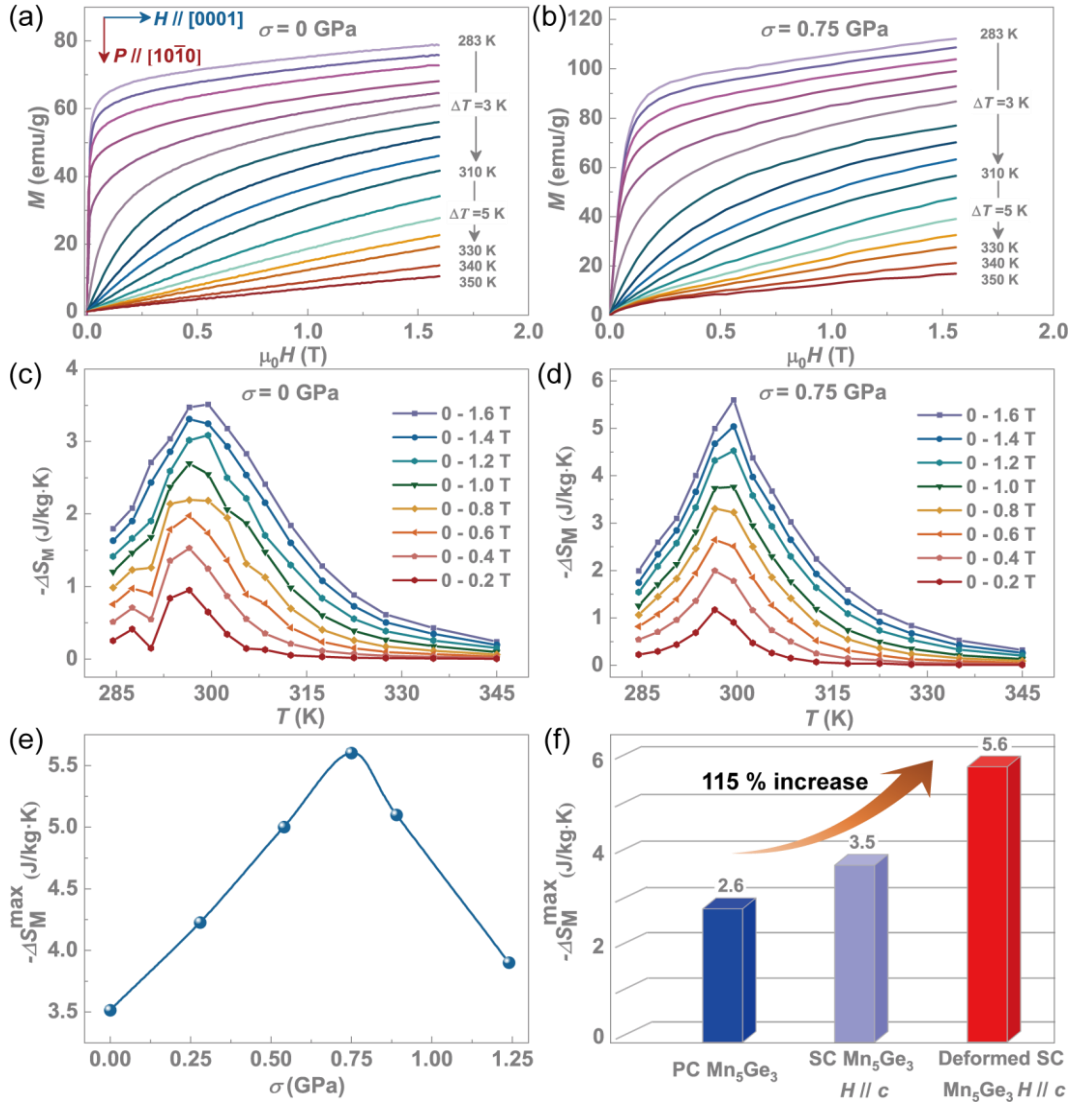

**Figure 3**. Enhanced magnetocaloric effect of Mn$_5$Ge$_3$ single crystals. (a, b) External magnetic field dependence of magnetization for the as-prepared sample ($\sigma$ = 0 GPa) and deformed sample ($\sigma$ = 0.75 GPa) at different temperatures. For all the measurements, $H$ parallels the [0001] direction. (c, d) Temperature dependence of magnetic entropy change $-\Delta S_M$ for the as-prepared sample ($\sigma$ = 0 GPa) and deformed sample ($\sigma$ = 0.75 GPa) under different external magnetic fields. (e) Stress dependence of maximum magnetic entropy change $-\Delta S_M^{max}$ for Mn$_5$Ge$_3$. (f) Comparison of $-\Delta S_M^{max}$ of the polycrystalline (PC) Mn$_5$Ge$_3$, the single-crystal (SC) Mn$_5$Ge$_3$ and the deformed SC Mn$_5$Ge$_3$.

Figure 3(e) plots the $\sigma$ dependence of $-\Delta S_M^{max}$ for Mn$_5$Ge$_3$. It can be seen that with the increase of $\sigma$, the magnitude of $-\Delta S_M^{max}$ does not monotonically increase but shows a peak shape and reaches a peak value of 5.6 J/kg·K at $\sigma$ = 0.75 GPa. This result has a significant 60 % improvement over that of $\sigma$ = 0 GPa at peak temperature of $T_{peak}$ = 300 K. This value is even 115 % higher than the magnetic entropy change of the polycrystalline sample [Figure 3(f)] and is comparable to Gd ($-\Delta S_M^{max}$ = 6.0 J/kg·K, $\Delta H$ = 2.0 T), which has been used as a benchmark material in magnetic refrigerator prototypes.[29] It is worth noting that the value of $-\Delta S_M^{max}$ does not show a monotonical increase trend with the increase of strain energy density [Figure 2(c)]. This may be due to a deterioration in magnetization caused by poly-crystallization and the appearance of other phases after deformation of $\sigma$ > 0.75 GPa, as discussed in the section of Figure 2.

To verify the accuracy of the above experimental results, we perform the field dependence of the magnetic entropy change. For SOMT materials, the field dependence of the $-\Delta S_M^{max}$ follows the power law[33,34]: $-\Delta S_M^{max} \propto H^n$, where $n$ is the power exponent associated with temperature and applied external magnetic field. The field dependence of magnetic entropy change has the following empirical features[33]: for temperatures well below Curie temperature ($T_C$), $n$ = 1; well above $T_C$, $n$ = 2; and at the peak temperature corresponding to $-\Delta S_M^{max}$, $n \approx 0.75$. In our case, the value of $n$ varies between 1 and 2 with temperature for all the samples before and after deformation [Figure S8]. Figure 4(a) shows the temperature dependence of $n$ for the as-prepared sample ($\sigma$ = 0 GPa) and deformed sample ($\sigma$ = 0.75 GPa). Insets show the fitting results by the power law at the peak temperature corresponding to $-\Delta S_M^{max}$ of $T_{pk}$ = 300 K. For the as-prepared sample, $n$ = 0.75 at $T_{pk}$ = 300 K. This value agrees with the above empirical features. For the deformed sample, $n$ = 0.83 at $T_{pk}$ = 300 K, slightly higher than 0.75. Considering that the value of $n$ at $T$ = 297 K is only 0.70, this result may be caused by rapid changes in $\Delta S_M$ over a very narrow temperature range.

Figure 4(b) shows the temperature dependence of magnetization with different stress ($\sigma$ = 0 and 0.75 GPa) of the Mn$_5$Ge$_3$ single crystals. The field cooling (FC) and zero-field cooling (ZFC) curves almost coincide, indicating no thermal hysteresis in our samples. The corresponding derivatives of the FC curves demonstrate that the Curie temperature of the as-prepared and deformed samples are 298 K and 299 K [inset of Figure 4(b)].

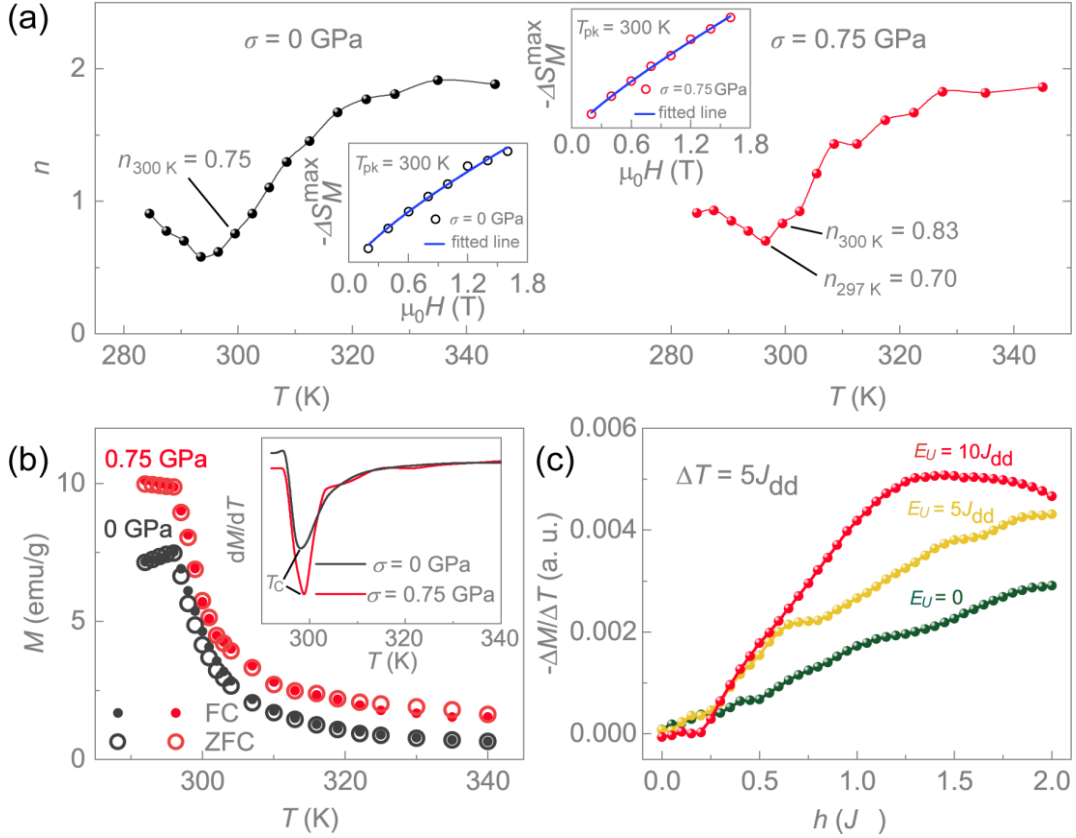

**Figure 4**. (a) Temperature dependence of the power exponent characterizing the field dependence of $-\Delta S_M$ for the as-prepared sample ($\sigma$ = 0 GPa) and deformed sample ($\sigma$ = 0.75 GPa). Insets show the power law behavior of maximum entropy change at $T_{pk}$ = 300 K for the as-prepared and deformed samples. (b) Temperature dependence of magnetization with different stress ($\sigma$ = 0 and 0.75 GPa) of the Mn$_5$Ge$_3$ single crystals. Solid dots and open circles represent the field cooling (FC) and zero-field cooling (ZFC) processes. For all the measurements, $H$ parallels the [0001] direction. The inset shows the corresponding derivatives of the FC $M$ – $T$ curves. (c) Simulated $-\Delta M/\Delta T$ of Mn$_5$Ge$_3$ with different $E_U$ = 0 $J_{dd}$, 5 $J_{dd}$, and 10 $J_{dd}$. $h$ is the external magnetic field reduced by $J_{dd}$.

To further explain the mechanism of strain energy enhancing magnetization in Mn$_5$Ge$_3$, we perform Monte Carlo simulations using the Ising-like Hamiltonian[26,35]

$$\mathcal{H} = -\sum_{i \neq j} J_{ij} \mathbf{S}_i \cdot \mathbf{S}_j - \mathbf{h} \cdot \sum_i g_i \mathbf{S}_i - E_U \quad (2)$$

where $J_{ij}$ is the exchange constant between the spins $\boldsymbol{S}_i$ and $\boldsymbol{S}_j$, $\boldsymbol{h}$ is the external magnetic field, $g_i$ is the Lande factor, and $E_U$ is the strain energy. Figure 4(c) shows the simulated $-\Delta M/\Delta T$ of Mn$_5$Ge$_3$ at $\Delta T = 5\ J_{dd}$. With increasing $E_U$, the value of $-\Delta M/\Delta T$ has an apparent tendency to increase. This result is consistent with the trend of our experimental results and strongly indicates that the $\Delta S_M$ of Mn$_5$Ge$_3$ can be intentionally tuned by the introduction of strain energy.

3. Conclusion

In summary, we realize a significant 60 % improvement in the magnetic entropy change at room temperature in Mn$_5$Ge$_3$ single crystals by introducing microscopic strain and lattice deformation with high-pressure constrained deformation. Both theoretical simulations and experimental results demonstrate that the prominent increase in magnetic entropy change originates from the tunable strain energy introduced by deformation. This additional strain energy will reconstruct the energy landscape of this ferromagnetic system and generate an enhancement in magnetization, resulting in a giant intensity of magnetocaloric responses. Our results provide an approach to increase magnetic entropy change and may give fresh ideas for exploring advanced magnetocaloric materials.

4. Experimental Section

*Materials synthesis*: Mn$_5$Ge$_3$ single crystals were synthesized using the bismuth-flux growth method. Mn (99.98 %), Ge (99.999 %), and Bi (99.99 %) were mixed in a stoichiometric ratio of 5:3:55 and placed in an alumina crucible sealed into a quartz tube under vacuum. The quartz tube was heated to 1050 °C in 10 hours and held for 72 hours. The temperature was slowly cooled to 650 °C at a rate of 2 °C/h. The Mn$_5$Ge$_3$ single crystals were then centrifugally separated from the bismuth flux using an embedded corundum filter. The composition stoichiometry was checked by energy-dispersive X-ray (EDX) spectroscopy combined with scanning electron microscopy (SEM) (S-4800, HITACHI).

*High-pressure constrained deformation*: The deformations were carried out using a commercial 600 kN machine. The power of this anisotropic severe plastic deformation technique has also been verified in constructing high-performance magnetic materials.[18,19] The process of high-pressure constrained deformation for the $Mn_5Ge_3$ single crystals is sketched in Figure S2. The hexagonal $Mn_5Ge_3$ single crystals were placed into the stacked steel wafers. Then, the steel wafers were sealed in a steel capsule. A press was applied to the assembled sample to realize the high-pressure constrained deformation. The press direction parallels the [10$\bar{1}$0] crystal orientation. The strength of deformation was expressed in terms of stress ($\sigma$) and the corresponding macro-strain ($\varepsilon$) along the direction of stress. Finally, the deformed sample was separated from the steel capsule and polished.

*Structural characterization*: The crystal structure of the single-crystal $Mn_5Ge_3$ was characterized by x-ray diffraction (XRD) with Co $K\alpha$ radiation ($\lambda$ = 1.789 Å) in a step-scanning mode (PIXcel-3D, PANalytical B.V., Inc.). The microstructure of samples was characterized by transmission electron microscopy (TEM) (JEM-2100 Plus, JEOL).

*Magnetization measurement*: Isothermal initial magnetization curves of samples were measured along the stress direction using a vibrating sample magnetometer (VSM) (Model-7404, 1.6 T, −10 °C ~ 500 °C, Lakeshore). The magnetic entropy change ($\Delta S_M$) was calculated by $\Delta S_M = \int_0^H (\partial M/\partial T)dH$.

*Monte Carlo simulation*: Monte Carlo simulations were performed using the Ising-like Hamiltonian $\mathcal{H} = -\sum_{i \neq j} J_{ij} \mathbf{S}_i \cdot \mathbf{S}_j - \mathbf{h} \cdot \sum_i g_i \mathbf{S}_i - E_U$, where $J_{ij}$ is the exchange constant between the spins $\mathbf{S}_i$ and $\mathbf{S}_j$, $\mathbf{h}$ is the external magnetic field, $g_i$ is the Lande factor, and $E_U$ is the strain energy. For Mn1 (4$d$ sites), $S = \pm 1.5$, while for Mn2 (6$g$ sites), $S = \pm 1.0$, and $g = 2.06$ for both the Mn1 and Mn2. For the exchange constant, we assign $J_{dd}$ for Mn1–Mn1, $J_{gg}$ for Mn2–Mn2, and $J_{dg}$ for Mn1–Mn2, respectively. During the simulation, we scale the exchange constants $J_{gg}$ and $J_{dg}$, the temperature ($T$), the external magnetic field ($h$), and the strain energy ($E_U$) to $J_{dd}$. We construct a 1000-cell superlattice with periodic boundary conditions and select $J_{dd}$ = 2.44 meV, $J_{gg}$ = 1.875 $J_{dd}$, and $J_{dg}$ = 1.221 $J_{dd}$.

**Supporting Information**

Supporting Information is available from the Wiley Online Library or from the author.

**Acknowledgments**

The authors would like to thank Dr. Jie Ren for his assistance in the Monte Carlo simulations. This work was supported by the National Natural Science Foundation of China (Grant Nos. 52101233 and 52071279), the Hebei Natural Science Foundation (Grant No. E2022203010), and the Innovation Capability Improvement Project of Hebei Province (Grant No. 22567605H).

**Conflict of Interest**

The authors declare no conflict of interest.

**Data Availability Statement**

The data that support the findings of this study are available from the corresponding author upon reasonable request.

**References**

[1] B. G. Shen, J. R. Sun, F. X. Hu, H. W. Zhang, Z. H. Cheng, *Adv. Mater.* **2009**, *21*, 4545.

[2] O. Gutfleisch, M. A. Willard, E. Brück, C. H. Chen, S. G. Sankar, J. P. Liu, *Adv. Mater.* **2011**, *23*, 821.

[3] J. Xiang, C. Zhang, Y. Gao, W. Schmidt, K. Schmalzl, C.-W. Wang, B. Li, N. Xi, X.-Y. Liu, H. Jin, G. Li, J. Shen, Z. Chen, Y. Qi, Y. Wan, W. Jin, W. Li, P. Sun, G. Su, *Nature* **2024**, *625*, 270.

[4] J. J. B. Levinsky, B. Beckmann, T. Gottschall, D. Koch, M. Ahmadi, O. Gutfleisch, G. R. Blake, *Nat. Commun.* **2024**, *15*, 8559.

[5] V. Franco, J. S. Blázquez, B. Ingale, A. Conde, *Annu. Rev. Mater. Res.* **2012**, *42*, 305.

[6] T. Gottschall, K. P. Skokov, M. Fries, A. Taubel, I. Radulov, F. Scheibel, D. Benke, S. Riegg, O. Gutfleisch, *Adv. Eng. Mater.* **2019**, *9*, 1901322.

[7] N. A. Zarkevich, V. I. Zverev, *Crystals* **2020**, *10*, 815.

[8] J. Y. Law, V. Franco, L. M. Moreno-Ramírez, A. Conde, D. Y. Karpenkov, I. Radulov, K. P. Skokov, O. Gutfleisch, *Nat. Commun.* **2018**, *9*, 2680.

[9] J. Y. Law, L. M. Moreno-Ramírez, Á. Díaz-García, V. Franco, *Journal of Applied Physics* **2023**, *133*, 040903.

[10] S. Singh, L. Caron, S. W. D'Souza, T. Fichtner, G. Porcari, S. Fabbrici, C. Shekhar, S. Chadov, M. Solzi, C. Felser, *Adv. Mater.* **2016**, *28*, 3321.

[11] S. Wang, C. Fan, D. Liu, *ACS Appl. Mater. Inter.* **2021**, *13*, 33237.

[12] L. Li, K. Nishimura, W. D. Hutchison, Z. Qian, D. Huo, T. NamiKi, *Appl. Phys. Lett.* **2012**, *100*, 152403.

[13] S. Xi, W. Lu, Y. Sun, *J. Appl. Phys.* **2012**, *111*, 063922.

[14] D. D. Belyea, M. S. Lucas, E. Michel, J. Horwath, C. W. Miller, *Sci. Rep.* **2015**, *5*, 15755.

[15] J. Z. Hao, F. X. Hu, H. B. Zhou, W. H. Liang, Z. B. Yu, F. R. Shen, Y. H. Gao, K. M. Qiao, J. Li, C. Zhang, B. J. Wang, J. Wang, J. He, J. R. Sun, B. G. Shen, *Scripta Mater.* **2020**, *186*, 84.

[16] J. Liu, *Chinese Phys. B* **2014**, *23*, 047503.

[17] H. Yin, J. Y. Law, Y. Huang, H. Shen, S. Jiang, S. Guo, V. Franco, J. Sun, *Sci. China Mater.* **2022**,

*65*, 1134.

[18] Y. Hua, X. Li, J. Li, X. Luo, Y. Li, W. Qin, L. Zhang, J. Xiao, W. Xia, P. Song, M. Yue, H. Zhang, X. Zhang, *Science* **2024**, *385*, 634.

[19] X. Li, L. Lou, W. Song, G. Huang, F. Hou, Q. Zhang, H. Zhang, J. Xiao, B. Wen, X. Zhang, *Adv. Mater.* **2017**, *29*, 1606430.

[20] L. Lou, Y. Li, X. Li, H. Li, W. Li, Y. Hua, W. Xia, Z. Zhao, H. Zhang, M. Yue, X. Zhang, *Adv. Mater.* **2021**, *33*, 2102800.

[21] X. Li, L. Lou, W. Song, Q. Zhang, G. Huang, Y. Hua, H. T. Zhang, J. Xiao, B. Wen, X. Zhang, *Nano Lett.* **2017**, *17*, 2985.

[22] W. Li, L. Li, Y. Nan, X. Li, X. Zhang, D. V. Gunderov, V. V. Stolyarov, A. G. Popov, *Appl. Phys. Lett.* **2007**, *91*, 062509.

[23] Y. Liu, L. Xu, Q. Wang, W. Li, X. Zhang, *Appl. Phys. Lett.* **2009**, *94*, 172502.

[24] H. Li, L. Lou, F. Hou, D. Guo, W. Li, X. Li, D. V. Gunderov, K. Sato, X. Zhang, *Appl. Phys. Lett.* **2013**, *103*, 142406.

[25] X. Zhang, *Mater. Res. Lett.* **2020**, *8*, 49.

[26] P. Song, S. Yao, B. Zhang, B. Jiang, S. Deng, D. Guo, L. Ma, D. Hou, *Appl. Phys. Lett.* **2022**, *120*, 192401.

[27] G. Huang, X. Li, L. Lou, Y. Hua, G. Zhu, M. Li, H. Zhang, J. Xiao, B. Wen, M. Yue, X. Zhang, *Small* **2018**, *14*, 1800619.

[28] S. Picozzi, A. Continenza, A. J. Freeman, *Phys. Rev. B* **2004**, *70*, 235205.

[29] K. A. Gschneidner, V. K. Pecharsky, *Annu. Rev. Mater. Sci.* **2000**, *30*, 387.

[30] M. Zhao, W. Guo, X. Wu, L. Ma, P. Song, G. Li, C. Zhen, D. Zhao, D. Hou, *Mater. Horiz.* **2023**, *10*, 4597.

[31] J. V. Selinger, *Introduction to the Theory of Soft Matter: From Ideal Gases to Liquid Crystals*, Springer International Publishing, Cham, **2016**.

[32] V. Biju, N. Sugathan, V. Vrinda, S. L. Salini, *J. Marer. Sci.* **2008**, *43*, 1175.

[33] V. Franco, J. S. Blázquez, A. Conde, *Appl. Phys. Lett.* **2006**, *89*, 222512.

[34] Lalita, P. D. Babu, Pardeep, G. A. Basheed, *J. Appl. Phys.* **2023**, *134*, 173903.

[35] A. Ghediri, Y. Chiba, A. Tlemçani, In *Artificial Intelligence and Heuristics for Smart Energy Efficiency in Smart Cities* (Ed.: Hatti, M.), Springer International Publishing, Cham, **2022**, pp. 847–856.